\documentclass[english,aps,prl,reprint]{revtex4-1}
\usepackage[T1]{fontenc}
\usepackage[latin9]{inputenc}
\usepackage{amsbsy}
\usepackage{amstext}
\usepackage{amssymb}
\usepackage{graphicx}
\usepackage{esint}
\usepackage{babel}
\begin{document}
\global\long\def\grad{\nabla}
\global\long\def\ap{A_{1\|}}
\global\long\def\apa{A_{\mathrm{\|ant}}}
\global\long\def\apt{A_{1\|\mathrm{tot}}}
\global\long\def\kps{k_{\perp}^{2}}
\global\long\def\p{\|}
\global\long\def\j{_{\sigma}}
\global\long\def\d{\mathrm{d}}
\global\long\def\jj{_{0\sigma}}
\global\long\def\sq{^{2}}
\global\long\def\r{_{\mathrm{ref}}}
\global\long\def\apb{\overline{A}_{1\sigma\|}}
\global\long\def\apab{\overline{A}_{\mathrm{ant}}}
\global\long\def\aptb{\overline{A}_{1\|\mathrm{tot},\sigma}}
\global\long\def\pb{\overline{\phi}_{1\sigma}}
\global\long\def\jb{J_{0}(\lambda_{\sigma})}
\global\long\def\apk{A_{1\|\boldsymbol{k}}}
\global\long\def\ppk{\phi_{1\boldsymbol{k}}}
\global\long\def\aptk{A_{1\|\mathrm{tot},\boldsymbol{k}}}
\global\long\def\apak{A_{\|\mathrm{ant},\boldsymbol{k}}}
\global\long\def\cc{\mathrm{c.c.}}
\global\long\def\fk{f_{\sigma\boldsymbol{k}}}
\global\long\def\gk{g_{\sigma\boldsymbol{k}}}
\global\long\def\apbk{\overline{A}_{1\|\boldsymbol{k}}}
\global\long\def\hk{h_{\sigma\boldsymbol{k}}}
\global\long\def\apabk{\overline{A}_{\mathrm{\|ant},\sigma\boldsymbol{k}}}
\global\long\def\aptbk{\overline{A}_{1\|\mathrm{tot},\sigma\boldsymbol{k}}}
\global\long\def\ppkb{\overline{\phi}_{1\sigma\boldsymbol{k}}}
\global\long\def\ki{k_{\perp}\rho_{i}}
\global\long\def\ke{k_{\perp}\rho_{e}}
\global\long\def\sp{\sigma}
\global\long\def\gene{\mathrm{GENE}}
\global\long\def\Alf{\text{{Alfvén}}}

\title{Multiscale nature of the dissipation range in gyrokinetic simulations
of Alfvénic turbulence}

\author{D.~Told}

\email{dtold@physics.ucla.edu}

\affiliation{Department of Physics and Astronomy, University of California, Los
Angeles, CA 90095, USA}

\affiliation{Max-Planck-Institut für Plasmaphysik, Boltzmannstraße 2, 85748 Garching,
Germany}

\author{F.~Jenko}

\affiliation{Department of Physics and Astronomy, University of California, Los
Angeles, CA 90095, USA}

\affiliation{Max-Planck-Institut für Plasmaphysik, Boltzmannstraße 2, 85748 Garching,
Germany}

\author{J.M.~TenBarge}

\affiliation{IREAP, University of Maryland, College Park, MD 20742, USA}

\author{G.G.~Howes}

\affiliation{Department of Physics and Astronomy, University of Iowa, Iowa City,
IA 52242, USA}

\author{G.W.~Hammett}

\affiliation{Princeton Plasma Physics Laboratory, Princeton, NJ 08543, USA}
\begin{abstract}
Nonlinear energy transfer and dissipation in Alfvén wave turbulence
are analyzed in the first gyrokinetic simulation spanning all scales
from the tail of the MHD range to the electron gyroradius scale. For
typical solar wind parameters at 1\,AU, about 30\% of the nonlinear
energy transfer close to the electron gyroradius scale is mediated
by modes in the tail of the MHD cascade. Collisional dissipation
occurs across the entire kinetic range $\ki\gtrsim1$. Both mechanisms
thus act on multiple coupled scales, which have to be retained for
a comprehensive picture of the dissipation range in Alfvénic turbulence.
\end{abstract}
\maketitle
\textit{Introduction.} Spacecraft measurements find a radial temperature
profile of the solar wind which can only be explained by the presence
of heating throughout the heliosphere \citep{Richardson03}. The key
mechanism of heating in the inner heliosphere up to $\sim$20\,AU
is thought to be the dissipation of turbulent fluctuation energy,
and its understanding and description is one of the outstanding open
issues in space physics \citep{Bruno13}. Over the past decade, numerous
studies, both observational \citep{Bale05,Alexandrova09,Sahraoui09,Sahraoui10,Chen13a}
and theoretical/computational \citep{Howes08,Howes11,Wan12,Salem12,TenBarge13,TenBarge2013,Osman14},
have focused on this topic, extracting ever more sophisticated measurements
of solar wind fluctuation properties, and accomplishing increasingly
detailed turbulence simulations. 

As the solar wind plasma is only weakly collisional, a variety of
kinetic effects such as cyclotron damping, Landau and transit time
damping, finite Larmor radius effects, stochastic heating, or particle
acceleration at reconnection sites can contribute to the conversion
of field energy to particle energy, and thus determine how collisional
dissipation will ultimately set in. A kinetic description is crucial
in order to judge the relative importance of each of those effects.
Due to the complexity of a nonlinear kinetic system, numerical simulations
are essential to interpret observations and provide guidance for analytical
theory. 

In the present Letter, we employ an approach based on gyrokinetic
(GK) theory \citep{Brizard07}, which is a rigorous limit of kinetic
theory in strongly magnetized plasmas. Due to the assumptions of low
frequencies (compared to the ion cyclotron frequency) and small fluctuation
levels, the gyrokinetic model excludes cyclotron resonances and stochastic
heating. In absence of these effects, we focus on the energetic properties
of kinetic $\Alf$ wave (KAW) turbulence, which has been demonstrated
to be a crucial ingredient of solar wind turbulence \citep{Podesta2013}.

We address the following key questions: (1) Which spectral features
can be found in a comprehensive simulation extending from the magnetohydrodynamic
(MHD) range down to the electron gyroradius scale? (2) What are the
characteristics of nonlinear energy transfer from large to small scales?
 (3) How is energy dissipated, and how is the dissipated energy partitioned
between ions and electrons? 

\textit{Simulation setup.} The nonlinear GK system of equations is
solved using the Eulerian code $\gene$ \citep{Jenko00} to study
the dynamics of KAW turbulence in three spatial dimensions. In order
to model the energy injection at the outer scales of the system, a
magnetic antenna potential, whose amplitude is evolved in time according
to a Langevin equation \citep{TenBarge14}, is externally prescribed
at the largest scales of the simulation domain. The driven modes are
$\left(0,1,\pm1\right)$ and $\left(1,0,\pm1\right)$, where $\left(i,j,k\right)$
are multiples of the lowest wave numbers in $\left(k_{x},k_{y},k_{z}\right)$,
respectively. The mean antenna frequency is chosen to be $\omega_{a}=0.9\omega_{A0}$
($\omega_{A0}$ being the frequency of the slowest $\Alf$ wave in
the system), the decorrelation rate is set to $\gamma_{a}=0.7\omega_{A0}$,
and the normalized antenna amplitude is set to $A_{\|,0}=\omega_{A0}B_{0}\sqrt{\delta}/C_{2}k_{\perp0}^{2}\sqrt{N}v_{A}$
(setting $\delta=2,$ $N=4$, $C_{2}=1$), in accordance with the
critical balance condition $\omega_{\mathrm{lin}}\sim\omega_{\mathrm{nl}}$
\citep{TenBarge14}. 

The physical parameters are chosen to be similar to solar wind conditions
at 1\,AU, with $\beta_{i}=8\pi n_{i}T_{i}/B_{0}^{2}=1$, $T_{i}/T_{e}=1$.
Proton and electron species are included with their real mass ratio
of $m_{i}/m_{e}=1836$. The electron collisionality is chosen to be
$\nu_{e}=0.06\omega_{A0}$ (with $\nu_{i}=\sqrt{m_{e}/m_{i}}\nu_{e}$),
a value small enough to not inhibit kinetic effects, but large enough
to reduce resolution requirements in velocity space.

In order to maximize the effective dynamic range, the simulation domain
is extended significantly compared to previously published work, to
include scales larger than the ion gyroradius, allowing for a free
distribution of energy into the KAW or the ion entropy cascade \citep{Schekochihin09}
as the ion gyroradius scale is passed. The evolution of the gyrocenter
distribution is tracked on a grid with the resolution $\left(n_{x},n_{y},n_{z},n_{v_{\|}},n_{\mu},n_{\sp}\right)=\left(512,512,96,48,15,2\right).$
The plane perpendicular to the background magnetic field is resolved
by $512^{2}$ fully dealiased grid points, covering a perpendicular
wavenumber range $0.2\le\ki\le51.2$ (or $0.0047\le k_{\perp}\rho_{e}\le1.19$),
thus extending into the regime where electron finite-Larmor-radius
effects become important. Here, $\rho_{\sp}=\sqrt{T_{\sp}m_{\sp}}c/eB$
with the species index $\sp$. The number of grid points in the perpendicular
plane is thus increased by a factor of 36 with respect to the largest
runs of this kind published to date \citep{Howes11}. 96 points are
used to resolve the dynamics along the background field (the $z$
direction), and $48\times15$ gridpoints are chosen to represent the
$\left(v_{\|},\mu\right)$ domain, where $v_{\|}$ is the velocity
along the guide field, and $\mu=mv_{\perp}^{2}/2B_{0}$ is the magnetic
moment with respect to the guide field. The domain sizes in velocity
space are chosen to extend up to 3 thermal velocities $v_{T\sp}$
in both parallel and perpendicular velocities for each species $\sp$,
where $v_{T\sp}=\sqrt{2T_{\sp}/m_{\sp}}$. 

Our simulations are performed using the same iterative expansion scheme
as in Ref.~\citep{Howes11}, where simulations are initially run
with low resolution and are then restarted several times with an increasingly
fine grid, until the target resolution is reached. The total runtime
is chosen to span several antenna oscillation periods $\tau_{A}$
(in this case $t_{\mathrm{end}}=7.20\tau_{A}$) in order to ensure
that a quasi-steady state has been reached. 

\emph{Diagnostic methods}. The key results of this study are obtained
using a set of sophisticated energy diagnostics (partially introduced
in Refs.~\citep{Banon2011,Banon2011a,Teaca12,Teaca2014}), which
enable studies of energy source, transfer and dissipation spectra
separately for each species, and which are applied to KAW turbulence
for the first time here. In particular, we analyze the time derivative
of the spatially averaged free energy density, which can be expressed
in the case of an antenna-driven electromagnetic system as
\begin{eqnarray}
\partial_{t}\mathcal{E} & = & \Re\sum_{\sp}\sum_{\boldsymbol{k}}\left\langle \frac{2\pi B_{0}}{m_{\sp}}\int\d\mu\d v_{\|}\left(\hk\frac{T_{0\sp}}{F_{0\sp}}\right.\right.\nonumber \\
 & + & \left.\frac{q_{\sp}v_{\|}}{c}\mathcal{C}\apabk\biggr)^{*}\partial_{t}\gk\right\rangle \nonumber \\
 & + & \Re\sum_{\boldsymbol{k}}\left\langle \frac{\kps}{4\pi}\aptk^{*}\partial_{t}\apak\right\rangle .\label{eq:EB}
\end{eqnarray}
Here, the sum over $\boldsymbol{k}$ denotes a summation over all
wavenumber pairs $\left(k_{x},k_{y}\right)$, and the angle brackets
indicate a spatial average along the guide field. $\fk$ is the perturbed
gyrocenter distribution, and $\hk=\fk+\left(q_{\sp}\ppkb+\mu\overline{B}_{1\|\sp\boldsymbol{k}}\right)F_{0\sp}/T_{0\sp}$
is its nonadiabatic part. The overbar denotes an average over the
gyro-ring, and $F_{0\sp}$ is a Maxwellian background distribution
with background density $n_{0\sp}$ and temperature $T_{0\sp}$. The
magnetic potential $\aptk=\apk+\apak$ is understood to contain also
the contribution due to the Langevin antenna $\apak$, which is necessary
for a complete account of the energy contained in the system. The
time derivative $\partial_{t}\gk=\partial_{t}(\fk+q_{\sp}v_{\|}\apbk F_{0\sp}/cT_{0\sp})$
is the quantity explicitly evolved in the GK Vlasov equation as implemented
in GENE, and 
\[
\mathcal{C}=\kps\biggl/\left(\kps+\sum_{\sp}\frac{8\pi^{2}q_{\sp}^{2}B_{0}}{m_{\sp}c^{2}T_{0\sp}}\int v_{\|}^{2}J_{0}^{2}\left(\lambda_{\sigma}\right)F_{0\sp}\d v_{\|}\d\mu\right)
\]
is a factor arising from the antenna-modified Ampere's law, with $\lambda_{\sp}=k_{\perp}\sqrt{2m_{\sp}\mu/B_{0}q_{\sp}^{2}}$.
By replacing $\partial_{t}\gk$ in Eq.~(\ref{eq:EB}) with any of
the various terms contributing to its evolution, we can assess the
impact of that term on the evolution of the free energy density. The
nonlinear transfer function (i.e. the free energy balance contribution
of the nonlinear term) thus reads 
\begin{eqnarray}
T_{\boldsymbol{kpq}} & = & \frac{\pi B_{0}}{m_{\sp}}\Re\int\mathrm{d}v_{\|}\mathrm{d}\mu\left[p_{x}q_{y}-p_{y}q_{x}\right]\left[\overline{\chi}_{1\sp\boldsymbol{p}}h_{\sp\boldsymbol{q}}-\overline{\chi}_{1\sp\boldsymbol{q}}h_{\sp\boldsymbol{p}}\right]\nonumber \\
 & \times & \left[\hk\frac{T_{0\sp}}{F_{0\sp}}+q_{\sp}v_{\|}\mathcal{C}\apabk/c\right],\label{eq:NLT}
\end{eqnarray}
with $\boldsymbol{k}+\boldsymbol{p}+\boldsymbol{q}=0$. Compared to
the definition used in Refs.~\citep{Teaca12,Teaca2014}, there is
an additional term involving the antenna potential, and the electrostatic
approximation has been dropped by using the full electromagnetic potential
$\overline{\chi}_{1\sp}=\overline{\phi}_{1\sp}-v_{\|}\aptb/c+\mu\overline{B}_{1\|\sp}/q_{\sp}$.
Note that the new antenna potential term does not satisfy the same
symmetry properties as the rest of the transfer function, consistent
with the fact that the antenna acts as an energy source through the
nonlinear term (but also through the parallel advection term). This
source can be quantified by measuring the symmetric part of the above
transfer function.

\begin{figure}
\includegraphics[width=1\columnwidth]{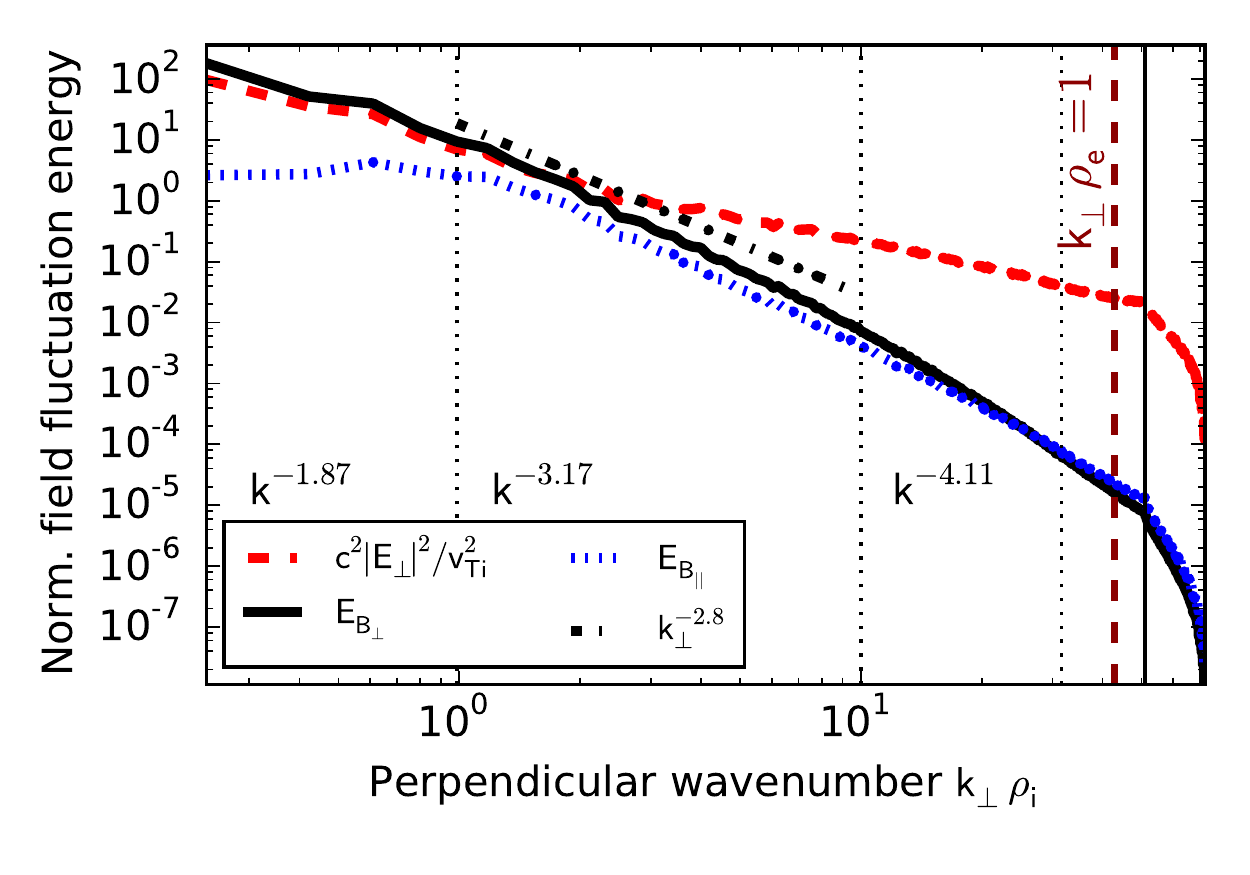}

\protect\caption{Normalized field energy spectra. \label{fig:Field-energy-spectra-b1}
Power law exponents obtained from the $B_{\perp}$ energy spectra
within the dotted sections are printed into the plot. }
\end{figure}

\textit{Field energy spectra.}  Before focusing on the nonlinear
transfer physics, we analyze the spectra of the magnetic and electric
field energy, which can be directly compared to spacecraft observations.
As is common practice, we compute 1-D spectra of $E_{E_{\perp}},$
$E_{B_{\|}}$, and $E_{B_{\perp}}$ vs. $k_{\perp}\rho_{i}$ by summing
the energy of all $\left(k_{x},k_{y}\right)$ modes within a given
$k_{\perp}$ shell. Shells are linearly spaced and divided into 384
bins; a short-time average over about $0.01\tau_{A}$ is performed,
as well as an average in $z$ direction. The results are displayed
in Fig.~\ref{fig:Field-energy-spectra-b1}. Here, the solid vertical
line denotes the boundary to the 'corner modes', for which the angle
integration in $\left(k_{x,}k_{y}\right)$ ceases to pick up complete
circles, causing the artificial spectral break. 

In the range $k_{\perp}\rho_{i}\lesssim1$, an MHD-type spectrum can
be observed, which exhibits a very small amount of compressive fluctuation
energy with a flat spectrum, and electric and magnetic field energy
spectra decaying approximately with the same power law. The power
law exponent is close to the Goldreich-Sridhar estimate of -5/3 \citep{GS95},
but the confidence level at small wavenumbers is low as there are
few modes per shell. 

As the range of $k_{\perp}\rho_{i}\sim1$ is crossed, all spectra
steepen, and the turbulence becomes more compressible (evidenced by
the increased ratio $\left|B_{\|}\right|^{2}/\left|B_{\perp}\right|^{2}$).
For $2\lesssim k_{\perp}\rho_{i}\lesssim15$, all quantities exhibit
rather well-defined power law spectra, until a further steepening
of the spectra sets in at $k_{\perp}\rho_{i}\approx15$, accompanied
by a crossing of the parallel and perpendicular magnetic fluctuation
energy. These spectral features are consistent with previous simulations
using a fraction of the present dynamic range \citep{TenBarge13}.
As the choice of parameters is (except for the collisionality) similar
to near-Earth solar wind measurements, in Fig.~\ref{fig:Field-energy-spectra-b1}
we plot the power law exponent $E_{B}\propto k_{\perp}^{-2.8}$ obtained
from the measurements of Refs.~\citep{Alexandrova09,Sahraoui13}
for comparison, which agrees within about 15\% with our average exponent
of -3.17, measured between $1<k_{\perp}\rho_{i}<10$. 

\emph{Nonlinear energy transfer.} In order to study the nonlinear
energy transfer, it is useful and necessary to reduce the data by
subdividing the perpendicular wavenumber plane into shells (see also
Ref.~\citep{Teaca12}), which we define as the region $0\leq k_{\perp}\leq k_{0}$
for the $0^{\mathrm{th}}$ shell and $k_{0}2^{(n-1)/3}\leq k_{\perp}\leq k_{0}2^{n/3}$
for the shells numbered $1\leq n\leq N-1$, where we set $k_{0}=0.275$
and $N=25$. Thus, the entire $k_{\perp}$ range present in the simulations
is covered, with good resolution also for $k_{\perp}\rho_{i}<1$,
while at the same time ensuring that only the lowest shell $0<k_{\perp}<k_{0}$
contains the externally driven modes.

\begin{figure}
\includegraphics[bb=10bp 25bp 323bp 277bp,clip,width=1\columnwidth]{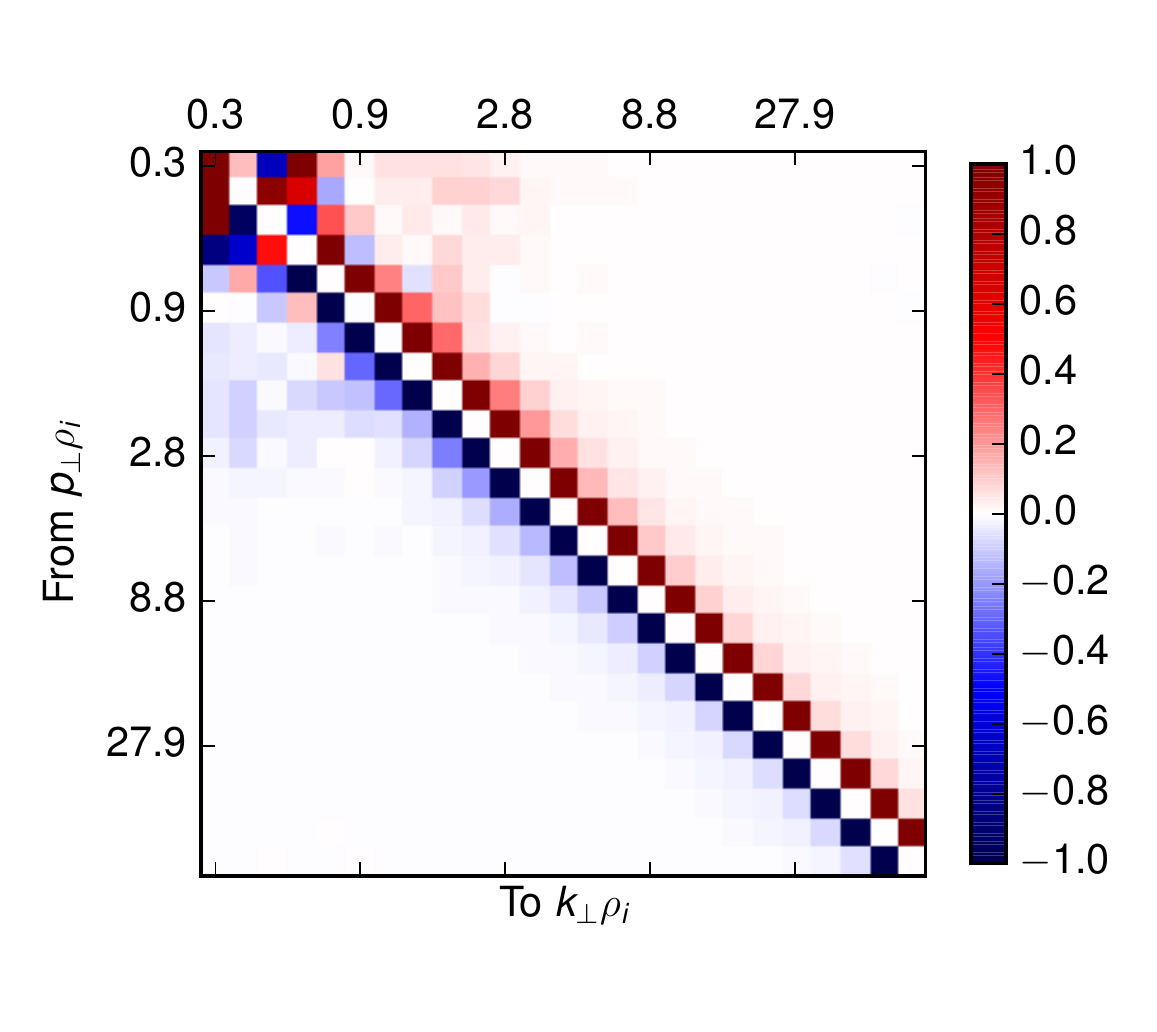}

\protect\caption{Nonlinear shell-to-shell transfer function for electrons, normalized
to the maximum absolute value of each wavenumber scale.\label{fig:NLT}}

\end{figure}
With this setup, we analyze the net nonlinear shell-to-shell energy
transfer, which is obtained by summing over all $q$ wavenumbers in
Eq.~(\ref{eq:NLT}). The resulting matrix (including the symmetric
terms due to the antenna, and normalized for each $k_{\perp}$ scale)
is displayed for the electron species in Fig.~\ref{fig:NLT}. Numerical
inspection shows that the antenna source acts almost exclusively on
the lowest shell, and diminishes very quickly for higher shell numbers.
Studying the conservative transfer more closely, one can observe that
in the range $\ki\lesssim3$, while local energy transfer dominates,
there are some nonlocal contributions connecting disparate $k_{\perp}$
scales. In the range $\ki>3$, on the other hand, the nonlinear transfer
is quite local $\left(k_{\perp}\approx p_{\perp}\right)$, i.e. dominated
by direct energy transfer between neighboring shells. 

\emph{Nonlocal mediation.} Beyond the net energy transfer, we now
extend the analysis to differentiate between different mediators,
i.e. $q$ wavenumbers. To this end, we evaluate the transfer function
of Eq.~(\ref{eq:NLT}) with triply filtered inputs, i.e., with fields
and distributions condensed into shells $K,P,Q$. Even with the limited
number of wavenumber shells used here, this diagnostic is extremely
expensive (approximately $\propto N^{2}$, or about 150,000 core-hours
here), and is thus only evaluated instantaneously for a single timestep.
Its results can be visualized in a compact way, e.g., by means of
Kraichnan's locality functions \citep{Kraichnan59}. The so-called
infrared (IR) locality function is defined (following the notation
of Ref.~\citep{Teaca12}) as 
\[
\Pi(k_{p}|k_{c})=\sum_{K=c+1}^{N}\left[\sum_{P=1}^{N}\sum_{Q=1}^{p}+\sum_{P=1}^{p}\sum_{Q=p+1}^{N}\right]T_{K,P,Q}
\]
and retains, for a fixed shell $k_{c}$ with a varying 'probe' wavenumber
$k_{p}$, only transfers for which at least one leg $p$ or $q$ is
smaller than $k_{p}$. Thus, starting with $k_{p}=k_{c}$ (retaining
all transfers) and then moving the probe $k_{p}$ away from $k_{c}$,
the most local transfers are successively removed. For an extensive
description of this setup, we refer the reader to Sec.~V of Ref.~\citep{Teaca2014}.
\begin{figure}
\includegraphics[bb=10bp 10bp 323bp 200bp,clip,width=1\columnwidth]{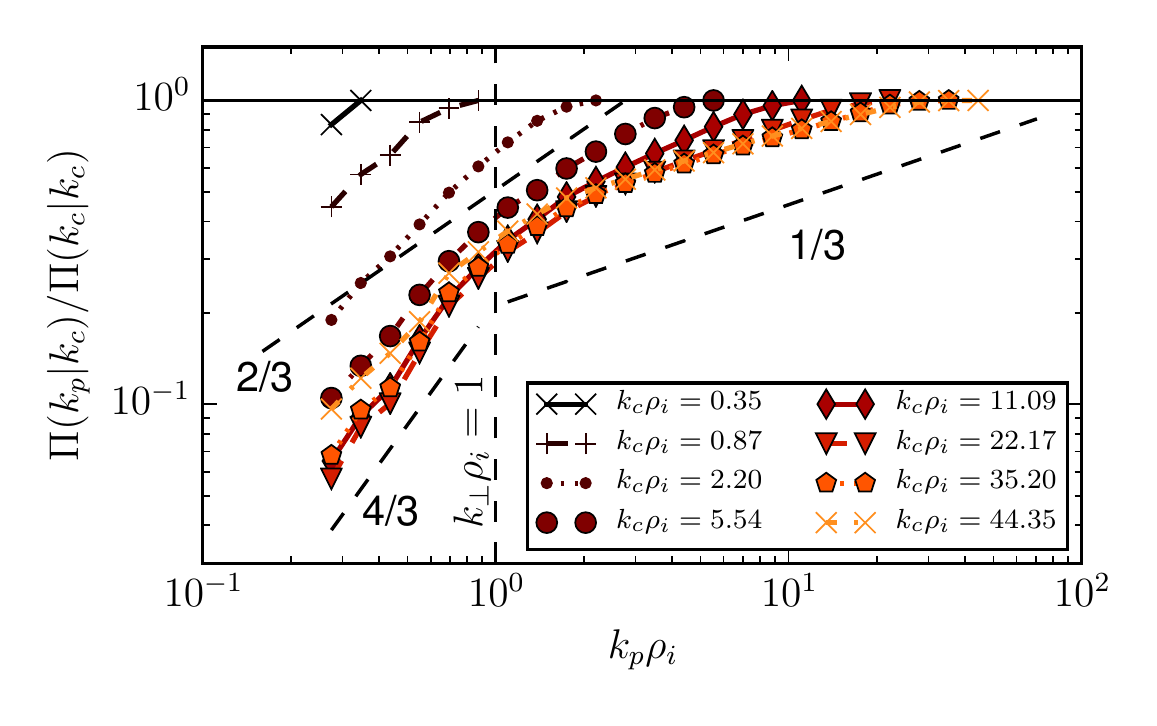}

\protect\caption{Infrared locality functions for several shells $k_{c}$, normalized
to the total nonlinear energy transfer through $k_{c}$, versus the
probe wavenumber $k_{p}\rho_{i}$. For the curves with $k_{c}\rho_{i}\gtrsim5$,
a change in slope is apparent when the probe $k_{p}$ crosses the
ion gyroradius scale. \label{fig:IR}}
\end{figure}

For several $k_{c}$ shells, we show the corresponding IR locality
functions $\Pi\left(k_{p}|k_{c}\right)/\Pi\left(k_{c}|k_{c}\right)$
in Fig.~\ref{fig:IR}. By plotting the curves versus the probe wavenumber
$k_{p}$ instead of the conventional ratio $k_{p}/k_{c}$, Fig.~\ref{fig:IR}
highlights the existence of a meaningful physical scale length at
$\ki\sim1$, indicating a lack of self-similarity. Indeed, the locality
function curves for $k_{c}\rho_{i}\gtrsim5$ exhibit a transition
in their slope that occurs close to the ion gyroradius scale, $k_{p}\rho_{i}\sim1$:
for $k_{p}\rho_{i}>1$ the nonlinear energy transfer is rather nonlocal,
with a locality exponent between 2/3 and 1/3; for $k_{p}\rho_{i}<1$,
a more local exponent of 4/3, as in Navier-Stokes turbulence \citep{Kraichnan66},
is found. As a consequence of this property, for $5\lesssim k_{c}\lesssim51.2$,
nonlocal transfers mediated by fluctuations in the tail of the MHD
range at $k_{p}\rho_{i}\lesssim1$ are responsible for at least 30\%
of the total energy transfer through these shells. Note that this
does not contradict the above observation that the \emph{net} nonlinear
transfer for large $k_{\perp}$ is local. Indeed, the nonlinear triad
$\boldsymbol{k}+\boldsymbol{p}+\boldsymbol{q}=0$ for such nonlocal
interactions is characterized by $\left|\boldsymbol{q}\right|\ll\left|\boldsymbol{k}\right|,\left|\boldsymbol{p}\right|$
and thus $\left|\boldsymbol{k}\right|\approx\left|\boldsymbol{p}\right|$,
consistent with a local net transfer between $\boldsymbol{k}$ and
\textbf{$\boldsymbol{p}$}. Finally, we note that while all of the
above statements were illustrated with results for the electron species,
the nonlinear ion energy transfer (not shown) exhibits the same characteristics,
though with an even more pronounced nonlocality (exponent $\sim1/12$),
and at least 50\% of the transfer mediated by modes in the tail of
the MHD range. 

\begin{figure}
\includegraphics[bb=10bp 10bp 318bp 188bp,clip,width=1\columnwidth]{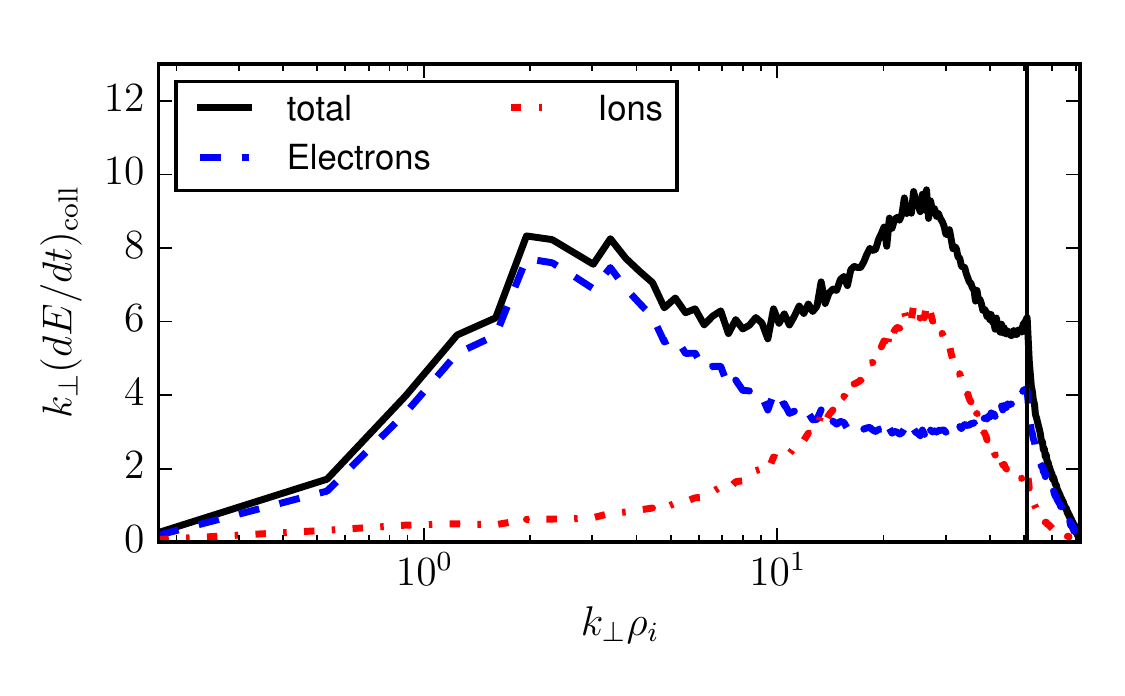}

\protect\caption{Normalized, short-time averaged collisional dissipation for electrons,
ions, and its total value. Curves are multiplied by $k_{\perp}$ so
the area under the curve is proportional to the energy dissipation
rate. \label{fig:colldiss}}
\end{figure}

\emph{Collisional dissipation.} Next, we study the spectral properties
of the collisional dissipation rate by measuring the contribution
of the collision term to the free energy balance. The resulting graphs
are presented in Fig.~\ref{fig:colldiss} for both electron and ion
species, as well as their sum. About 70\% of the total dissipation
is found to arise from electron collisions, which exhibit a broad
peak around $\ki\sim1-5$. Qualitatively, this peak is consistent
with electron Landau damping acting on the magnetic energy spectrum
shown in Figure~\ref{fig:Field-energy-spectra-b1}. Despite peaking
at these relatively small $k_{\perp}$ wavenumbers, electron dissipation
remains strong throughout the spectrum, and begins to intensify somewhat
at $\ki\gtrsim30$. At $\ki\sim1$, where ion transit-time damping
is expected to transfer field energy to ion particle energy, there
is in fact little ion heating. At these scales the ion free energy
(not shown) is comparable to the magnetic fluctuation energy, but
it is cascaded to smaller scales in both position and velocity space,
and is dissipated close to the \emph{electron} gyroradius scale (around
$\ki\sim25$). This observation is consistent with an ion entropy
cascade and the fact that $\nu_{i}\ll\nu_{e}$ \citep{Schekochihin09,Tatsuno09,Howes11}.
Taking into account both species' contributions, we find an essentially
flat dissipation spectrum \emph{throughout} the kinetic wavenumber
range, contrasting with some interpretations of solar wind data \citep{Alexandrova09,Sahraoui09}
which suggested that the electron gyroradius scale acts as the dominant
dissipation scale. 

\textit{Conclusions. }\textit{\emph{In the present study, the first
gyrokinetic simulation of kinetic $\Alf$ wave turbulence coupling
all scales from the tail of the MHD range to the electron gyroradius
scale was performed, with the goal of analyzing fundamental properties
of nonlinear energy transfer and collisional dissipation for parameters
relevant to the solar wind. It was found that nonlinear energy transfer
in the kinetic range, particularly for $\ki\gtrsim5$, is considerably
more nonlocal than hydrodynamic turbulence, as suggested by previous
theoretical considerations \citep{Howes11a}, and is to a significant
percentage ($>$30\%) mediated by the tail of the MHD cascade just
below $\ki\sim1$, while the net energy transfer occurs mainly between
nearest-neighbor shells. For $T_{e}/T_{i}=1$ and $\beta_{i}=1$,
similar to the near-Earth solar wind, 70\% of the injected energy
is dissipated through the electron species, whose dissipation spectrum
peaks around $\ki\sim1-5$, consistent with electron Landau damping.
The ion free energy, on the other hand, is cascaded to small scales
and dissipated around $\ki\sim25$. These findings underscore the
presence of strong dissipation throughout the kinetic range $\ki\gtrsim1$,
justifying the common notion of a 'dissipation range', and demonstrating
a coupling across multiple scales of both transfer and dissipation.}}

\emph{Acknowledgments}. The authors acknowledge fruitful discussions
with F.~Muller, A.~Bañón Navarro and M.J.~Pueschel. The research
leading to these results has received funding from the European Research
Council under the European Union's Seventh Framework Programme (FP7/2007-2013)/ERC
Grant Agreement No. 277870, NSF CAREER Award AGS-1054061, and U.S.
DOE Award No. DEFG0293ER54197. Furthermore, this work was facilitated
by the Max-Planck/Princeton Center for Plasma Physics. The Rechenzentrum
Garching (RZG) is gratefully acknowledged for providing computational
resources used for this study. Parts of this research also profited
from resources of the National Energy Research Scientific Computing
Center, a DOE Office of Science User Facility supported by the Office
of Science of the U.S. Department of Energy under Contract No. DE-AC02-05CH11231.

\bibliographystyle{apsrev4-1}

\begin{thebibliography}{29}%
\makeatletter
\providecommand \@ifxundefined [1]{%
 \@ifx{#1\undefined}
}%
\providecommand \@ifnum [1]{%
 \ifnum #1\expandafter \@firstoftwo
 \else \expandafter \@secondoftwo
 \fi
}%
\providecommand \@ifx [1]{%
 \ifx #1\expandafter \@firstoftwo
 \else \expandafter \@secondoftwo
 \fi
}%
\providecommand \natexlab [1]{#1}%
\providecommand \enquote  [1]{``#1''}%
\providecommand \bibnamefont  [1]{#1}%
\providecommand \bibfnamefont [1]{#1}%
\providecommand \citenamefont [1]{#1}%
\providecommand \href@noop [0]{\@secondoftwo}%
\providecommand \href [0]{\begingroup \@sanitize@url \@href}%
\providecommand \@href[1]{\@@startlink{#1}\@@href}%
\providecommand \@@href[1]{\endgroup#1\@@endlink}%
\providecommand \@sanitize@url [0]{\catcode `\\12\catcode `\$12\catcode
  `\&12\catcode `\#12\catcode `\^12\catcode `\_12\catcode `\%12\relax}%
\providecommand \@@startlink[1]{}%
\providecommand \@@endlink[0]{}%
\providecommand \url  [0]{\begingroup\@sanitize@url \@url }%
\providecommand \@url [1]{\endgroup\@href {#1}{\urlprefix }}%
\providecommand \urlprefix  [0]{URL }%
\providecommand \Eprint [0]{\href }%
\providecommand \doibase [0]{http://dx.doi.org/}%
\providecommand \selectlanguage [0]{\@gobble}%
\providecommand \bibinfo  [0]{\@secondoftwo}%
\providecommand \bibfield  [0]{\@secondoftwo}%
\providecommand \translation [1]{[#1]}%
\providecommand \BibitemOpen [0]{}%
\providecommand \bibitemStop [0]{}%
\providecommand \bibitemNoStop [0]{.\EOS\space}%
\providecommand \EOS [0]{\spacefactor3000\relax}%
\providecommand \BibitemShut  [1]{\csname bibitem#1\endcsname}%
\let\auto@bib@innerbib\@empty
\bibitem [{\citenamefont {{Richardson}}\ and\ \citenamefont
  {{Smith}}(2003)}]{Richardson03}%
  \BibitemOpen
  \bibfield  {author} {\bibinfo {author} {\bibfnamefont {J.~D.}\ \bibnamefont
  {{Richardson}}}\ and\ \bibinfo {author} {\bibfnamefont {C.~W.}\ \bibnamefont
  {{Smith}}},\ }\href {\doibase 10.1029/2002GL016551} {\bibfield  {journal}
  {\bibinfo  {journal} {Geophys.~Res.~Lett.}\ }\textbf {\bibinfo {volume}
  {30}},\ \bibinfo {eid} {1206} (\bibinfo {year} {2003})}\BibitemShut {NoStop}%
\bibitem [{\citenamefont {{Bruno}}\ and\ \citenamefont
  {{Carbone}}(2013)}]{Bruno13}%
  \BibitemOpen
  \bibfield  {author} {\bibinfo {author} {\bibfnamefont {R.}~\bibnamefont
  {{Bruno}}}\ and\ \bibinfo {author} {\bibfnamefont {V.}~\bibnamefont
  {{Carbone}}},\ }\href {\doibase 10.12942/lrsp-2013-2} {\bibfield  {journal}
  {\bibinfo  {journal} {Living~Rev.~Sol.~Phys.}\ }\textbf {\bibinfo {volume}
  {10}},\ \bibinfo {pages} {2} (\bibinfo {year} {2013})}\BibitemShut {NoStop}%
\bibitem [{\citenamefont {{Bale}}\ \emph {et~al.}(2005)\citenamefont {{Bale}},
  \citenamefont {{Kellogg}}, \citenamefont {{Mozer}}, \citenamefont
  {{Horbury}},\ and\ \citenamefont {{Reme}}}]{Bale05}%
  \BibitemOpen
  \bibfield  {author} {\bibinfo {author} {\bibfnamefont {S.~D.}\ \bibnamefont
  {{Bale}}}, \bibinfo {author} {\bibfnamefont {P.~J.}\ \bibnamefont
  {{Kellogg}}}, \bibinfo {author} {\bibfnamefont {F.~S.}\ \bibnamefont
  {{Mozer}}}, \bibinfo {author} {\bibfnamefont {T.~S.}\ \bibnamefont
  {{Horbury}}}, \ and\ \bibinfo {author} {\bibfnamefont {H.}~\bibnamefont
  {{Reme}}},\ }\href {\doibase 10.1103/PhysRevLett.94.215002} {\bibfield
  {journal} {\bibinfo  {journal} {Phys.~Rev.~Lett.}\ }\textbf {\bibinfo
  {volume} {94}},\ \bibinfo {eid} {215002} (\bibinfo {year} {2005})},\ \Eprint
  {http://arxiv.org/abs/physics/0503103} {physics/0503103} \BibitemShut
  {NoStop}%
\bibitem [{\citenamefont {{Alexandrova}}\ \emph {et~al.}(2009)\citenamefont
  {{Alexandrova}}, \citenamefont {{Saur}}, \citenamefont {{Lacombe}},
  \citenamefont {{Mangeney}}, \citenamefont {{Mitchell}}, \citenamefont
  {{Schwartz}},\ and\ \citenamefont {{Robert}}}]{Alexandrova09}%
  \BibitemOpen
  \bibfield  {author} {\bibinfo {author} {\bibfnamefont {O.}~\bibnamefont
  {{Alexandrova}}}, \bibinfo {author} {\bibfnamefont {J.}~\bibnamefont
  {{Saur}}}, \bibinfo {author} {\bibfnamefont {C.}~\bibnamefont {{Lacombe}}},
  \bibinfo {author} {\bibfnamefont {A.}~\bibnamefont {{Mangeney}}}, \bibinfo
  {author} {\bibfnamefont {J.}~\bibnamefont {{Mitchell}}}, \bibinfo {author}
  {\bibfnamefont {S.~J.}\ \bibnamefont {{Schwartz}}}, \ and\ \bibinfo {author}
  {\bibfnamefont {P.}~\bibnamefont {{Robert}}},\ }\href {\doibase
  10.1103/PhysRevLett.103.165003} {\bibfield  {journal} {\bibinfo  {journal}
  {Phys.~Rev.~Lett.}\ }\textbf {\bibinfo {volume} {103}},\ \bibinfo {eid}
  {165003} (\bibinfo {year} {2009})},\ \Eprint {http://arxiv.org/abs/0906.3236}
  {arXiv:0906.3236 [physics.plasm-ph]} \BibitemShut {NoStop}%
\bibitem [{\citenamefont {{Sahraoui}}\ \emph {et~al.}(2009)\citenamefont
  {{Sahraoui}}, \citenamefont {{Goldstein}}, \citenamefont {{Robert}},\ and\
  \citenamefont {{Khotyaintsev}}}]{Sahraoui09}%
  \BibitemOpen
  \bibfield  {author} {\bibinfo {author} {\bibfnamefont {F.}~\bibnamefont
  {{Sahraoui}}}, \bibinfo {author} {\bibfnamefont {M.~L.}\ \bibnamefont
  {{Goldstein}}}, \bibinfo {author} {\bibfnamefont {P.}~\bibnamefont
  {{Robert}}}, \ and\ \bibinfo {author} {\bibfnamefont {Y.~V.}\ \bibnamefont
  {{Khotyaintsev}}},\ }\href {\doibase 10.1103/PhysRevLett.102.231102}
  {\bibfield  {journal} {\bibinfo  {journal} {Phys.~Rev.~Lett.}\ }\textbf
  {\bibinfo {volume} {102}},\ \bibinfo {eid} {231102} (\bibinfo {year}
  {2009})}\BibitemShut {NoStop}%
\bibitem [{\citenamefont {{Sahraoui}}\ \emph {et~al.}(2010)\citenamefont
  {{Sahraoui}}, \citenamefont {{Goldstein}}, \citenamefont {{Belmont}},
  \citenamefont {{Canu}},\ and\ \citenamefont {{Rezeau}}}]{Sahraoui10}%
  \BibitemOpen
  \bibfield  {author} {\bibinfo {author} {\bibfnamefont {F.}~\bibnamefont
  {{Sahraoui}}}, \bibinfo {author} {\bibfnamefont {M.~L.}\ \bibnamefont
  {{Goldstein}}}, \bibinfo {author} {\bibfnamefont {G.}~\bibnamefont
  {{Belmont}}}, \bibinfo {author} {\bibfnamefont {P.}~\bibnamefont {{Canu}}}, \
  and\ \bibinfo {author} {\bibfnamefont {L.}~\bibnamefont {{Rezeau}}},\ }\href
  {\doibase 10.1103/PhysRevLett.105.131101} {\bibfield  {journal} {\bibinfo
  {journal} {Phys.~Rev.~Lett.}\ }\textbf {\bibinfo {volume} {105}},\ \bibinfo
  {eid} {131101} (\bibinfo {year} {2010})}\BibitemShut {NoStop}%
\bibitem [{\citenamefont {{Chen}}\ \emph {et~al.}(2013)\citenamefont {{Chen}},
  \citenamefont {{Boldyrev}}, \citenamefont {{Xia}},\ and\ \citenamefont
  {{Perez}}}]{Chen13a}%
  \BibitemOpen
  \bibfield  {author} {\bibinfo {author} {\bibfnamefont {C.~H.~K.}\
  \bibnamefont {{Chen}}}, \bibinfo {author} {\bibfnamefont {S.}~\bibnamefont
  {{Boldyrev}}}, \bibinfo {author} {\bibfnamefont {Q.}~\bibnamefont {{Xia}}}, \
  and\ \bibinfo {author} {\bibfnamefont {J.~C.}\ \bibnamefont {{Perez}}},\
  }\href {\doibase 10.1103/PhysRevLett.110.225002} {\bibfield  {journal}
  {\bibinfo  {journal} {Phys.~Rev.~Lett.}\ }\textbf {\bibinfo {volume} {110}},\
  \bibinfo {eid} {225002} (\bibinfo {year} {2013})},\ \Eprint
  {http://arxiv.org/abs/1305.2950} {arXiv:1305.2950 [physics.space-ph]}
  \BibitemShut {NoStop}%
\bibitem [{\citenamefont {{Howes}}\ \emph {et~al.}(2008)\citenamefont
  {{Howes}}, \citenamefont {{Dorland}}, \citenamefont {{Cowley}}, \citenamefont
  {{Hammett}}, \citenamefont {{Quataert}}, \citenamefont {{Schekochihin}},\
  and\ \citenamefont {{Tatsuno}}}]{Howes08}%
  \BibitemOpen
  \bibfield  {author} {\bibinfo {author} {\bibfnamefont {G.~G.}\ \bibnamefont
  {{Howes}}}, \bibinfo {author} {\bibfnamefont {W.}~\bibnamefont {{Dorland}}},
  \bibinfo {author} {\bibfnamefont {S.~C.}\ \bibnamefont {{Cowley}}}, \bibinfo
  {author} {\bibfnamefont {G.~W.}\ \bibnamefont {{Hammett}}}, \bibinfo {author}
  {\bibfnamefont {E.}~\bibnamefont {{Quataert}}}, \bibinfo {author}
  {\bibfnamefont {A.~A.}\ \bibnamefont {{Schekochihin}}}, \ and\ \bibinfo
  {author} {\bibfnamefont {T.}~\bibnamefont {{Tatsuno}}},\ }\href {\doibase
  10.1103/PhysRevLett.100.065004} {\bibfield  {journal} {\bibinfo  {journal}
  {Phys.~Rev.~Lett.}\ }\textbf {\bibinfo {volume} {100}},\ \bibinfo {eid}
  {065004} (\bibinfo {year} {2008})},\ \Eprint {http://arxiv.org/abs/0711.4355}
  {arXiv:0711.4355} \BibitemShut {NoStop}%
\bibitem [{\citenamefont {{Howes}}\ \emph
  {et~al.}(2011{\natexlab{a}})\citenamefont {{Howes}}, \citenamefont
  {{TenBarge}}, \citenamefont {{Dorland}}, \citenamefont {{Quataert}},
  \citenamefont {{Schekochihin}}, \citenamefont {{Numata}},\ and\ \citenamefont
  {{Tatsuno}}}]{Howes11}%
  \BibitemOpen
  \bibfield  {author} {\bibinfo {author} {\bibfnamefont {G.~G.}\ \bibnamefont
  {{Howes}}}, \bibinfo {author} {\bibfnamefont {J.~M.}\ \bibnamefont
  {{TenBarge}}}, \bibinfo {author} {\bibfnamefont {W.}~\bibnamefont
  {{Dorland}}}, \bibinfo {author} {\bibfnamefont {E.}~\bibnamefont
  {{Quataert}}}, \bibinfo {author} {\bibfnamefont {A.~A.}\ \bibnamefont
  {{Schekochihin}}}, \bibinfo {author} {\bibfnamefont {R.}~\bibnamefont
  {{Numata}}}, \ and\ \bibinfo {author} {\bibfnamefont {T.}~\bibnamefont
  {{Tatsuno}}},\ }\href {\doibase 10.1103/PhysRevLett.107.035004} {\bibfield
  {journal} {\bibinfo  {journal} {Phys.~Rev.~Lett.}\ }\textbf {\bibinfo
  {volume} {107}},\ \bibinfo {eid} {035004} (\bibinfo {year}
  {2011}{\natexlab{a}})},\ \Eprint {http://arxiv.org/abs/1104.0877}
  {arXiv:1104.0877 [astro-ph.SR]} \BibitemShut {NoStop}%
\bibitem [{\citenamefont {{Wan}}\ \emph {et~al.}(2012)\citenamefont {{Wan}},
  \citenamefont {{Matthaeus}}, \citenamefont {{Karimabadi}}, \citenamefont
  {{Roytershteyn}}, \citenamefont {{Shay}}, \citenamefont {{Wu}}, \citenamefont
  {{Daughton}}, \citenamefont {{Loring}},\ and\ \citenamefont
  {{Chapman}}}]{Wan12}%
  \BibitemOpen
  \bibfield  {author} {\bibinfo {author} {\bibfnamefont {M.}~\bibnamefont
  {{Wan}}}, \bibinfo {author} {\bibfnamefont {W.~H.}\ \bibnamefont
  {{Matthaeus}}}, \bibinfo {author} {\bibfnamefont {H.}~\bibnamefont
  {{Karimabadi}}}, \bibinfo {author} {\bibfnamefont {V.}~\bibnamefont
  {{Roytershteyn}}}, \bibinfo {author} {\bibfnamefont {M.}~\bibnamefont
  {{Shay}}}, \bibinfo {author} {\bibfnamefont {P.}~\bibnamefont {{Wu}}},
  \bibinfo {author} {\bibfnamefont {W.}~\bibnamefont {{Daughton}}}, \bibinfo
  {author} {\bibfnamefont {B.}~\bibnamefont {{Loring}}}, \ and\ \bibinfo
  {author} {\bibfnamefont {S.~C.}\ \bibnamefont {{Chapman}}},\ }\href {\doibase
  10.1103/PhysRevLett.109.195001} {\bibfield  {journal} {\bibinfo  {journal}
  {Phys.~Rev.~Lett.}\ }\textbf {\bibinfo {volume} {109}},\ \bibinfo {eid}
  {195001} (\bibinfo {year} {2012})}\BibitemShut {NoStop}%
\bibitem [{\citenamefont {{Salem}}\ \emph {et~al.}(2012)\citenamefont
  {{Salem}}, \citenamefont {{Howes}}, \citenamefont {{Sundkvist}},
  \citenamefont {{Bale}}, \citenamefont {{Chaston}}, \citenamefont {{Chen}},\
  and\ \citenamefont {{Mozer}}}]{Salem12}%
  \BibitemOpen
  \bibfield  {author} {\bibinfo {author} {\bibfnamefont {C.~S.}\ \bibnamefont
  {{Salem}}}, \bibinfo {author} {\bibfnamefont {G.~G.}\ \bibnamefont
  {{Howes}}}, \bibinfo {author} {\bibfnamefont {D.}~\bibnamefont
  {{Sundkvist}}}, \bibinfo {author} {\bibfnamefont {S.~D.}\ \bibnamefont
  {{Bale}}}, \bibinfo {author} {\bibfnamefont {C.~C.}\ \bibnamefont
  {{Chaston}}}, \bibinfo {author} {\bibfnamefont {C.~H.~K.}\ \bibnamefont
  {{Chen}}}, \ and\ \bibinfo {author} {\bibfnamefont {F.~S.}\ \bibnamefont
  {{Mozer}}},\ }\href {\doibase 10.1088/2041-8205/745/1/L9} {\bibfield
  {journal} {\bibinfo  {journal} {Astrophys. J. Lett.}\ }\textbf {\bibinfo
  {volume} {745}},\ \bibinfo {eid} {L9} (\bibinfo {year} {2012})}\BibitemShut
  {NoStop}%
\bibitem [{\citenamefont {{TenBarge}}\ \emph {et~al.}(2013)\citenamefont
  {{TenBarge}}, \citenamefont {{Howes}},\ and\ \citenamefont
  {{Dorland}}}]{TenBarge13}%
  \BibitemOpen
  \bibfield  {author} {\bibinfo {author} {\bibfnamefont {J.~M.}\ \bibnamefont
  {{TenBarge}}}, \bibinfo {author} {\bibfnamefont {G.~G.}\ \bibnamefont
  {{Howes}}}, \ and\ \bibinfo {author} {\bibfnamefont {W.}~\bibnamefont
  {{Dorland}}},\ }\href {\doibase 10.1088/0004-637X/774/2/139} {\bibfield
  {journal} {\bibinfo  {journal} {Astrophys. J.}\ }\textbf {\bibinfo {volume}
  {774}},\ \bibinfo {eid} {139} (\bibinfo {year} {2013})}\BibitemShut {NoStop}%
\bibitem [{\citenamefont {{TenBarge}}\ and\ \citenamefont
  {{Howes}}(2013)}]{TenBarge2013}%
  \BibitemOpen
  \bibfield  {author} {\bibinfo {author} {\bibfnamefont {J.~M.}\ \bibnamefont
  {{TenBarge}}}\ and\ \bibinfo {author} {\bibfnamefont {G.~G.}\ \bibnamefont
  {{Howes}}},\ }\href {\doibase 10.1088/2041-8205/771/2/L27} {\bibfield
  {journal} {\bibinfo  {journal} {Astrophys. J. Lett.}\ }\textbf {\bibinfo
  {volume} {771}},\ \bibinfo {eid} {L27} (\bibinfo {year} {2013})},\ \Eprint
  {http://arxiv.org/abs/1304.2958} {arXiv:1304.2958 [physics.plasm-ph]}
  \BibitemShut {NoStop}%
\bibitem [{\citenamefont {{Osman}}\ \emph {et~al.}(2014)\citenamefont
  {{Osman}}, \citenamefont {{Matthaeus}}, \citenamefont {{Gosling}},
  \citenamefont {{Greco}}, \citenamefont {{Servidio}}, \citenamefont {{Hnat}},
  \citenamefont {{Chapman}},\ and\ \citenamefont {{Phan}}}]{Osman14}%
  \BibitemOpen
  \bibfield  {author} {\bibinfo {author} {\bibfnamefont {K.~T.}\ \bibnamefont
  {{Osman}}}, \bibinfo {author} {\bibfnamefont {W.~H.}\ \bibnamefont
  {{Matthaeus}}}, \bibinfo {author} {\bibfnamefont {J.~T.}\ \bibnamefont
  {{Gosling}}}, \bibinfo {author} {\bibfnamefont {A.}~\bibnamefont {{Greco}}},
  \bibinfo {author} {\bibfnamefont {S.}~\bibnamefont {{Servidio}}}, \bibinfo
  {author} {\bibfnamefont {B.}~\bibnamefont {{Hnat}}}, \bibinfo {author}
  {\bibfnamefont {S.~C.}\ \bibnamefont {{Chapman}}}, \ and\ \bibinfo {author}
  {\bibfnamefont {T.~D.}\ \bibnamefont {{Phan}}},\ }\href {\doibase
  10.1103/PhysRevLett.112.215002} {\bibfield  {journal} {\bibinfo  {journal}
  {Phys.~Rev.~Lett.}\ }\textbf {\bibinfo {volume} {112}},\ \bibinfo {eid}
  {215002} (\bibinfo {year} {2014})},\ \Eprint {http://arxiv.org/abs/1403.4590}
  {arXiv:1403.4590 [physics.space-ph]} \BibitemShut {NoStop}%
\bibitem [{\citenamefont {Brizard}\ and\ \citenamefont
  {Hahm}(2007)}]{Brizard07}%
  \BibitemOpen
  \bibfield  {author} {\bibinfo {author} {\bibfnamefont {A.~J.}\ \bibnamefont
  {Brizard}}\ and\ \bibinfo {author} {\bibfnamefont {T.~S.}\ \bibnamefont
  {Hahm}},\ }\href {\doibase 10.1103/RevModPhys.79.421} {\bibfield  {journal}
  {\bibinfo  {journal} {Rev.~Mod.~Phys.}\ }\textbf {\bibinfo {volume} {79}},\
  \bibinfo {eid} {421} (\bibinfo {year} {2007})}\BibitemShut {NoStop}%
\bibitem [{\citenamefont {Podesta}(2013)}]{Podesta2013}%
  \BibitemOpen
  \bibfield  {author} {\bibinfo {author} {\bibfnamefont {J.~J.}\ \bibnamefont
  {Podesta}},\ }\href {\doibase 10.1007/s11207-013-0258-z} {\bibfield
  {journal} {\bibinfo  {journal} {Sol.~Phys.}\ }\textbf {\bibinfo {volume}
  {286}},\ \bibinfo {pages} {529} (\bibinfo {year} {2013})}\BibitemShut
  {NoStop}%
\bibitem [{\citenamefont {{Jenko}}\ \emph {et~al.}(2000)\citenamefont
  {{Jenko}}, \citenamefont {{Dorland}}, \citenamefont {{Kotschenreuther}},\
  and\ \citenamefont {{Rogers}}}]{Jenko00}%
  \BibitemOpen
  \bibfield  {author} {\bibinfo {author} {\bibfnamefont {F.}~\bibnamefont
  {{Jenko}}}, \bibinfo {author} {\bibfnamefont {W.}~\bibnamefont {{Dorland}}},
  \bibinfo {author} {\bibfnamefont {M.}~\bibnamefont {{Kotschenreuther}}}, \
  and\ \bibinfo {author} {\bibfnamefont {B.~N.}\ \bibnamefont {{Rogers}}},\
  }\href {\doibase 10.1063/1.874014} {\bibfield  {journal} {\bibinfo  {journal}
  {Phys.~Plasmas.}\ }\textbf {\bibinfo {volume} {7}},\ \bibinfo {pages} {1904}
  (\bibinfo {year} {2000})}\BibitemShut {NoStop}%
\bibitem [{\citenamefont {{TenBarge}}\ \emph {et~al.}(2014)\citenamefont
  {{TenBarge}}, \citenamefont {{Howes}}, \citenamefont {{Dorland}},\ and\
  \citenamefont {{Hammett}}}]{TenBarge14}%
  \BibitemOpen
  \bibfield  {author} {\bibinfo {author} {\bibfnamefont {J.~M.}\ \bibnamefont
  {{TenBarge}}}, \bibinfo {author} {\bibfnamefont {G.~G.}\ \bibnamefont
  {{Howes}}}, \bibinfo {author} {\bibfnamefont {W.}~\bibnamefont {{Dorland}}},
  \ and\ \bibinfo {author} {\bibfnamefont {G.~W.}\ \bibnamefont {{Hammett}}},\
  }\href {\doibase 10.1016/j.cpc.2013.10.022} {\bibfield  {journal} {\bibinfo
  {journal} {Comput.~Phys.~Commun.}\ }\textbf {\bibinfo {volume} {185}},\
  \bibinfo {pages} {578} (\bibinfo {year} {2014})},\ \Eprint
  {http://arxiv.org/abs/1305.2212} {arXiv:1305.2212 [physics.plasm-ph]}
  \BibitemShut {NoStop}%
\bibitem [{\citenamefont {{Schekochihin}}\ \emph {et~al.}(2009)\citenamefont
  {{Schekochihin}}, \citenamefont {{Cowley}}, \citenamefont {{Dorland}},
  \citenamefont {{Hammett}}, \citenamefont {{Howes}}, \citenamefont
  {{Quataert}},\ and\ \citenamefont {{Tatsuno}}}]{Schekochihin09}%
  \BibitemOpen
  \bibfield  {author} {\bibinfo {author} {\bibfnamefont {A.~A.}\ \bibnamefont
  {{Schekochihin}}}, \bibinfo {author} {\bibfnamefont {S.~C.}\ \bibnamefont
  {{Cowley}}}, \bibinfo {author} {\bibfnamefont {W.}~\bibnamefont {{Dorland}}},
  \bibinfo {author} {\bibfnamefont {G.~W.}\ \bibnamefont {{Hammett}}}, \bibinfo
  {author} {\bibfnamefont {G.~G.}\ \bibnamefont {{Howes}}}, \bibinfo {author}
  {\bibfnamefont {E.}~\bibnamefont {{Quataert}}}, \ and\ \bibinfo {author}
  {\bibfnamefont {T.}~\bibnamefont {{Tatsuno}}},\ }\href {\doibase
  10.1088/0067-0049/182/1/310} {\bibfield  {journal} {\bibinfo  {journal}
  {Astrophys. J. Suppl. Ser.}\ }\textbf {\bibinfo {volume} {182}},\ \bibinfo
  {pages} {310} (\bibinfo {year} {2009})},\ \Eprint
  {http://arxiv.org/abs/0704.0044} {arXiv:0704.0044} \BibitemShut {NoStop}%
\bibitem [{\citenamefont {{Ba{\~n}{\'o}n Navarro}}\ \emph
  {et~al.}(2011{\natexlab{a}})\citenamefont {{Ba{\~n}{\'o}n Navarro}},
  \citenamefont {{Morel}}, \citenamefont {{Albrecht-Marc}}, \citenamefont
  {{Carati}}, \citenamefont {{Merz}}, \citenamefont {{G{\"o}rler}},\ and\
  \citenamefont {{Jenko}}}]{Banon2011}%
  \BibitemOpen
  \bibfield  {author} {\bibinfo {author} {\bibfnamefont {A.}~\bibnamefont
  {{Ba{\~n}{\'o}n Navarro}}}, \bibinfo {author} {\bibfnamefont
  {P.}~\bibnamefont {{Morel}}}, \bibinfo {author} {\bibfnamefont
  {M.}~\bibnamefont {{Albrecht-Marc}}}, \bibinfo {author} {\bibfnamefont
  {D.}~\bibnamefont {{Carati}}}, \bibinfo {author} {\bibfnamefont
  {F.}~\bibnamefont {{Merz}}}, \bibinfo {author} {\bibfnamefont
  {T.}~\bibnamefont {{G{\"o}rler}}}, \ and\ \bibinfo {author} {\bibfnamefont
  {F.}~\bibnamefont {{Jenko}}},\ }\href {\doibase
  10.1103/PhysRevLett.106.055001} {\bibfield  {journal} {\bibinfo  {journal}
  {Phys.~Rev.~Lett.}\ }\textbf {\bibinfo {volume} {106}},\ \bibinfo {eid}
  {055001} (\bibinfo {year} {2011}{\natexlab{a}})},\ \Eprint
  {http://arxiv.org/abs/1008.3974} {arXiv:1008.3974 [physics.plasm-ph]}
  \BibitemShut {NoStop}%
\bibitem [{\citenamefont {{Ba{\~n}{\'o}n Navarro}}\ \emph
  {et~al.}(2011{\natexlab{b}})\citenamefont {{Ba{\~n}{\'o}n Navarro}},
  \citenamefont {{Morel}}, \citenamefont {{Albrecht-Marc}}, \citenamefont
  {{Carati}}, \citenamefont {{Merz}}, \citenamefont {{G{\"o}rler}},\ and\
  \citenamefont {{Jenko}}}]{Banon2011a}%
  \BibitemOpen
  \bibfield  {author} {\bibinfo {author} {\bibfnamefont {A.}~\bibnamefont
  {{Ba{\~n}{\'o}n Navarro}}}, \bibinfo {author} {\bibfnamefont
  {P.}~\bibnamefont {{Morel}}}, \bibinfo {author} {\bibfnamefont
  {M.}~\bibnamefont {{Albrecht-Marc}}}, \bibinfo {author} {\bibfnamefont
  {D.}~\bibnamefont {{Carati}}}, \bibinfo {author} {\bibfnamefont
  {F.}~\bibnamefont {{Merz}}}, \bibinfo {author} {\bibfnamefont
  {T.}~\bibnamefont {{G{\"o}rler}}}, \ and\ \bibinfo {author} {\bibfnamefont
  {F.}~\bibnamefont {{Jenko}}},\ }\href {\doibase 10.1063/1.3632077} {\bibfield
   {journal} {\bibinfo  {journal} {Phys.~Plasmas.}\ }\textbf {\bibinfo {volume}
  {18}},\ \bibinfo {eid} {092303} (\bibinfo {year}
  {2011}{\natexlab{b}})}\BibitemShut {NoStop}%
\bibitem [{\citenamefont {{Teaca}}\ \emph {et~al.}(2012)\citenamefont
  {{Teaca}}, \citenamefont {{Navarro}}, \citenamefont {{Jenko}}, \citenamefont
  {{Brunner}},\ and\ \citenamefont {{Villard}}}]{Teaca12}%
  \BibitemOpen
  \bibfield  {author} {\bibinfo {author} {\bibfnamefont {B.}~\bibnamefont
  {{Teaca}}}, \bibinfo {author} {\bibfnamefont {A.~B.}\ \bibnamefont
  {{Navarro}}}, \bibinfo {author} {\bibfnamefont {F.}~\bibnamefont {{Jenko}}},
  \bibinfo {author} {\bibfnamefont {S.}~\bibnamefont {{Brunner}}}, \ and\
  \bibinfo {author} {\bibfnamefont {L.}~\bibnamefont {{Villard}}},\ }\href
  {\doibase 10.1103/PhysRevLett.109.235003} {\bibfield  {journal} {\bibinfo
  {journal} {Phys.~Rev.~Lett.}\ }\textbf {\bibinfo {volume} {109}},\ \bibinfo
  {eid} {235003} (\bibinfo {year} {2012})}\BibitemShut {NoStop}%
\bibitem [{\citenamefont {{Teaca}}\ \emph {et~al.}(2014)\citenamefont
  {{Teaca}}, \citenamefont {{Navarro}},\ and\ \citenamefont
  {{Jenko}}}]{Teaca2014}%
  \BibitemOpen
  \bibfield  {author} {\bibinfo {author} {\bibfnamefont {B.}~\bibnamefont
  {{Teaca}}}, \bibinfo {author} {\bibfnamefont {A.~B.}\ \bibnamefont
  {{Navarro}}}, \ and\ \bibinfo {author} {\bibfnamefont {F.}~\bibnamefont
  {{Jenko}}},\ }\href {\doibase 10.1063/1.4890127} {\bibfield  {journal}
  {\bibinfo  {journal} {Phys.~Plasmas.}\ }\textbf {\bibinfo {volume} {21}},\
  \bibinfo {eid} {072308} (\bibinfo {year} {2014})},\ \Eprint
  {http://arxiv.org/abs/1404.2080} {arXiv:1404.2080 [physics.plasm-ph]}
  \BibitemShut {NoStop}%
\bibitem [{\citenamefont {{Goldreich}}\ and\ \citenamefont
  {{Sridhar}}(1995)}]{GS95}%
  \BibitemOpen
  \bibfield  {author} {\bibinfo {author} {\bibfnamefont {P.}~\bibnamefont
  {{Goldreich}}}\ and\ \bibinfo {author} {\bibfnamefont {S.}~\bibnamefont
  {{Sridhar}}},\ }\href {\doibase 10.1086/175121} {\bibfield  {journal}
  {\bibinfo  {journal} {Astrophys. J.}\ }\textbf {\bibinfo {volume} {438}},\
  \bibinfo {pages} {763} (\bibinfo {year} {1995})}\BibitemShut {NoStop}%
\bibitem [{\citenamefont {Sahraoui}\ \emph {et~al.}(2013)\citenamefont
  {Sahraoui}, \citenamefont {Huang}, \citenamefont {Belmont}, \citenamefont
  {Goldstein}, \citenamefont {R{\'e}tino}, \citenamefont {Robert},\ and\
  \citenamefont {Patoul}}]{Sahraoui13}%
  \BibitemOpen
  \bibfield  {author} {\bibinfo {author} {\bibfnamefont {F.}~\bibnamefont
  {Sahraoui}}, \bibinfo {author} {\bibfnamefont {S.~Y.}\ \bibnamefont {Huang}},
  \bibinfo {author} {\bibfnamefont {G.}~\bibnamefont {Belmont}}, \bibinfo
  {author} {\bibfnamefont {M.~L.}\ \bibnamefont {Goldstein}}, \bibinfo {author}
  {\bibfnamefont {A.}~\bibnamefont {R{\'e}tino}}, \bibinfo {author}
  {\bibfnamefont {P.}~\bibnamefont {Robert}}, \ and\ \bibinfo {author}
  {\bibfnamefont {J.~D.}\ \bibnamefont {Patoul}},\ }\href
  {http://stacks.iop.org/0004-637X/777/i=1/a=15} {\bibfield  {journal}
  {\bibinfo  {journal} {Astrophys.~J.}\ }\textbf {\bibinfo {volume} {777}},\
  \bibinfo {pages} {15} (\bibinfo {year} {2013})}\BibitemShut {NoStop}%
\bibitem [{\citenamefont {{Kraichnan}}(1959)}]{Kraichnan59}%
  \BibitemOpen
  \bibfield  {author} {\bibinfo {author} {\bibfnamefont {R.~H.}\ \bibnamefont
  {{Kraichnan}}},\ }\href {\doibase 10.1017/S0022112059000362} {\bibfield
  {journal} {\bibinfo  {journal} {J.~Fluid~Mech.}\ }\textbf {\bibinfo {volume}
  {5}},\ \bibinfo {pages} {497} (\bibinfo {year} {1959})}\BibitemShut {NoStop}%
\bibitem [{\citenamefont {{Kraichnan}}(1966)}]{Kraichnan66}%
  \BibitemOpen
  \bibfield  {author} {\bibinfo {author} {\bibfnamefont {R.~H.}\ \bibnamefont
  {{Kraichnan}}},\ }\href {\doibase 10.1063/1.1761928} {\bibfield  {journal}
  {\bibinfo  {journal} {Phys.~Fluids}\ }\textbf {\bibinfo {volume} {9}},\
  \bibinfo {pages} {1728} (\bibinfo {year} {1966})}\BibitemShut {NoStop}%
\bibitem [{\citenamefont {{Tatsuno}}\ \emph {et~al.}(2009)\citenamefont
  {{Tatsuno}}, \citenamefont {{Dorland}}, \citenamefont {{Schekochihin}},
  \citenamefont {{Plunk}}, \citenamefont {{Barnes}}, \citenamefont {{Cowley}},\
  and\ \citenamefont {{Howes}}}]{Tatsuno09}%
  \BibitemOpen
  \bibfield  {author} {\bibinfo {author} {\bibfnamefont {T.}~\bibnamefont
  {{Tatsuno}}}, \bibinfo {author} {\bibfnamefont {W.}~\bibnamefont
  {{Dorland}}}, \bibinfo {author} {\bibfnamefont {A.~A.}\ \bibnamefont
  {{Schekochihin}}}, \bibinfo {author} {\bibfnamefont {G.~G.}\ \bibnamefont
  {{Plunk}}}, \bibinfo {author} {\bibfnamefont {M.}~\bibnamefont {{Barnes}}},
  \bibinfo {author} {\bibfnamefont {S.~C.}\ \bibnamefont {{Cowley}}}, \ and\
  \bibinfo {author} {\bibfnamefont {G.~G.}\ \bibnamefont {{Howes}}},\ }\href
  {\doibase 10.1103/PhysRevLett.103.015003} {\bibfield  {journal} {\bibinfo
  {journal} {Phys.~Rev.~Lett.}\ }\textbf {\bibinfo {volume} {103}},\ \bibinfo
  {eid} {015003} (\bibinfo {year} {2009})},\ \Eprint
  {http://arxiv.org/abs/0811.2538} {arXiv:0811.2538 [physics.plasm-ph]}
  \BibitemShut {NoStop}%
\bibitem [{\citenamefont {{Howes}}\ \emph
  {et~al.}(2011{\natexlab{b}})\citenamefont {{Howes}}, \citenamefont
  {{TenBarge}},\ and\ \citenamefont {{Dorland}}}]{Howes11a}%
  \BibitemOpen
  \bibfield  {author} {\bibinfo {author} {\bibfnamefont {G.~G.}\ \bibnamefont
  {{Howes}}}, \bibinfo {author} {\bibfnamefont {J.~M.}\ \bibnamefont
  {{TenBarge}}}, \ and\ \bibinfo {author} {\bibfnamefont {W.}~\bibnamefont
  {{Dorland}}},\ }\href {\doibase 10.1063/1.3646400} {\bibfield  {journal}
  {\bibinfo  {journal} {Phys.~Plasmas.}\ }\textbf {\bibinfo {volume} {18}},\
  \bibinfo {pages} {102305} (\bibinfo {year} {2011}{\natexlab{b}})},\ \Eprint
  {http://arxiv.org/abs/1109.4158} {arXiv:1109.4158 [astro-ph.SR]} \BibitemShut
  {NoStop}%
\end{thebibliography}

\end{document}